\documentclass[showpacs,prl,preprint]{revtex4}

\usepackage{amssymb}
\usepackage{graphicx}

\begin{document}

\title{Electron spin relaxation in carbon nanotubes}
\author{Y. G. Semenov, J. M. Zavada, and K. W. Kim}

\begin{abstract}
The long standing problem of inexplicably short spin relaxation in
carbon nanotubes (CNTs) is examined. The curvature-mediated
spin-orbital interaction is shown to induce fluctuating electron
spin precession causing efficient relaxation in a manner analogous
to the Dyakonov-Perel mechanism. Our calculation estimates
longitudinal (spin-flip) and transversal (decoherence) relaxation
times as short as 150 ps and 110 ps at room temperature,
respectively, along with a pronounced anisotropic dependence.
Interference of electrons originating from different valleys can
lead to even faster dephasing. The results can help clarify the
measured data, resolving discrepancies in the literature.
\end{abstract}

\pacs{85.35.Kt, 85.75.-d, 81.07.Vb, 75.76.+j}

\address{Department of Electrical and Computer Engineering, North Carolina
State University, Raleigh, NC 27695-7911}

\maketitle

Spin-related effects in carbon nanotubes (CNTs) have attracted much
attention in recent years \cite{Cottet06}. Initially, it was assumed
that a weak spin-orbital interaction (SOI) in CNTs, due to
$p_{z}$-electrons, would lead to a long spin relaxation (and spin
coherence length) compared with typical semiconductors. However,
numerous experimental measurements of spin transport and electron
spin resonance in CNTs present very diparate results for the
magnitude of the spin relaxation time $\tau _{s}$. The processing
possibilities for spin transport impose spin conservation at least
for the dwell time $\tau _{d}$ of the electrons in CNT \cite{Fert07}
providing a lower limit for $\tau _{s}$. Such approach, however,
does not guarantee the direct derivation of $\tau _{s}$, moreover,
association of $\tau _{d}$ with time of electron drift between
ferromagnetic contacts leads to enormous discrepancy with spin
relaxation time measured by other methods. For example, a drift time
around a ps attributes to magnetoresistance (MR) measurements in
Refs. \onlinecite{Tsukagoshi99} and \onlinecite{Sahoo05}, while the
electron spin resonance indicates much longer $\tau _{s}\approx 5$
ns at temperature $T=300$ K \cite{Petit97}. Sufficient progress in
the analysis of spin transport in CNT has been
provided in Ref. \onlinecite{Hueso07} where both $\tau _{d}=60$ ns and $%
\tau _{s}=30$ ns were extracted from MR measurements at $T\leq 120$
K.

A theoretical approach to spin relaxation based on hyperfine
interaction \cite{SKI} failed to explain the short spin relaxation
in CNT, even through a recent study \cite{Churchill09N} corrected
the hyperfine interaction constant by two orders of magnitude, still
making the spin relaxation longer than 100 $\mu $s. While the carbon
intra-atomic SOI has been evaluated as 12 meV \cite{Cardona00},
higher orders of perturbation theory result in extremely weak
spin-orbital coupling ($\sim 1\mu $eV) for planar graphene
\cite{Brataas06}. Thus there is a negligible effect on spin
relaxation in CNTs.

Recently such an assumption was revised theoretically \cite%
{Ando00,Brataas06,Izumida09,Chico09,Jeong09} and experimentally~\cite%
{Kuemmeth08} because graphene curvature in CNTs produces mixing of $\pi $
and $\sigma $ electron states that can drastically enhance SOI.
Qualitatively, it can be viewed as electron spin in a magnetic field stemed
from a clockwise or counterclockwise circular motion around the CNT's
circumference. States with opposite electron motion originate from the two
different valleys $K$ and $K^{\prime }$ \cite{Ando00}. The finding of so
strong SOI was shown to be responsible for an efficient mechanism of
electron spin-lattice relaxation in CNT quantum dots at low temperatures
\cite{Bulaev08}. The actual mechanism of itinerate electron spin relaxation
in CNTs at room temperature still presents a very intriguing problem for
carbon based spintronics.

We begin with a few essential definitions concerning CNTs \cite{Ando05}. The
CNT is conveniently imagined as a spiral graphite sheet (graphene) rolled
along the chiral vector $\mathbf{C}_{h}$ (see Fig. 1). Here $\mathbf{C}%
_{h}=n_{a}\mathbf{a}+n_{b}\mathbf{b}$, where $\mathbf{a}=a(1,0)$ and $%
\mathbf{b}=a(1/2,\sqrt{3}/2)$ are the graphene lattice unit vectors
with $a=0.246$ nm and $n_{a}$ and $n_{b}$ are integers, which
characterize the geometry of a particular CNT. The slope of
$\mathbf{C}_{h}$ is defined by $\tan \theta
=\sqrt{3}n_{b}/(2n_{a}+n_{b})$ and the diameter of the CNT is given
by $d=\left\vert \mathbf{C}_{h}\right\vert /\pi $. The energy band
structure of graphene possesses two non-equivalent valleys with
Dirac-like dispersion law in the vicinity of Fermi level \cite{Slonczewski57}%
. They are located at the $K=\frac{2\pi }{3a}\left( 1,\sqrt{3}\right) $ and $%
K^{\prime }=\frac{2\pi }{3a}\left( -1,\sqrt{3}\right) $ corner
points of the first Brillouin zone, which will be labeled by
$\lambda =1,-1$ respectively.

The graphene two-valley band structure
projects onto the CNT one so that in the vicinity of each valley the
$kP$ Hamiltonian takes the form $H_{\lambda
}=\hbar v_{F}(\lambda \varkappa _{m}\widehat{\sigma }_{1}+k\widehat{\sigma }%
_{2})$ \cite{Ando05} where the Fermi velocity is $v_{F}=8\cdot 10^{7}$ cm/s
and Pauli \ matrixes $\widehat{\sigma }_{1}$ and $\widehat{\sigma }_{2}$ are
defined over the sublattice electronic states A and B. The wave vector $%
\mathbf{k}$ with respect to $K$ ($\lambda =1$) or $K^{\prime }$ ($\lambda
=-1 $) point is directed along the principal axis $\zeta $ of CNT. The
rolling along the perpendicular direction $\rho $ (Fig.~1) imposes a
quantization of the electron momentum in circumcircular direction fixing the
wave numbers $\varkappa _{m}=2(m-\lambda \nu /3)/d$ with integer magnetic
quantum number $m $ and $\nu $ (from the set 1, 0, -1) so that $%
2n_{a}+n_{b}+\nu $ becomes divisible by 3. If the CNT is subjected to a
magnetic field $\mathbf{B}_{0}$ the Aharonov-Bohm magnetic flux passing
through the CNT cross section $\phi _{AB}=B_{0}\cos \alpha \ \pi d^{2}/4$
modifies the quantization condition \cite{Ando05},
\begin{equation}
\varkappa _{m,\lambda }=2(m-\lambda \nu /3+ \phi _{AB}/\phi _{0})/d
, \label{2a}
\end{equation}%
where $\alpha $ is an angle between $\mathbf{B}_{0}$ and CNT axis, $\phi
_{0}=2\pi \hbar c/\left\vert q_{0}\right\vert $ is the flux quantum, $q_{0}$
the electron charge.

The eigenvalues of Hamiltonian $H_{\lambda }$ describe the CNT electronic
spectrum without spin structure,%
\begin{equation}
\varepsilon _{k,m,\lambda }=\pm \hbar v_{F}\sqrt{\varkappa _{m,\lambda
}^{2}+k^{2}},  \label{2b}
\end{equation}%
where + (-) corresponds to the conduction (valence) band. The CNTs with $\nu
=\pm 1$ possess the semiconducting spectrum, with bandgap $E_{g}=4\hbar
v_{F}/3d$ even at zero magnetic field, with potential applications in
spintronics \cite{Cottet06}. We assume $\nu =1$ in further numerical
calculations.

When a spin-orbital interaction is incorporated in the Hamiltonian $%
H_{\lambda }$, the SOI mediated by CNT curvature which takes the simplest
form in CNT co-ordinates $\xi $, $\eta $, $\zeta $ (Fig.~1) as:%
\begin{equation}
H_{SO}=\Delta _{0}\lambda \widehat{\sigma }_{0}2S_{\zeta }+\Delta
_{1}\lambda \widehat{\sigma }_{1}2S_{\zeta }-i\Delta _{2}\widehat{\sigma }%
_{2}(S_{+}e^{i\varphi }-S_{-}e^{-i\varphi }),  \label{3}
\end{equation}%
where $\widehat{\sigma }_{0}$ is $2\times 2$ identical matrix, $S_{\pm
}=S_{\xi }\pm iS_{\eta }$, $\varphi =2\rho /d$, $0<\rho <\left\vert \mathbf{C%
}_{h}\right\vert $. In~(\ref{3}) the spin-orbital constants are proportional
to CNT curvature, $\Delta _{0}=(\delta _{0}/d)\cos 3\theta $ , $\Delta
_{1,2}=\delta _{1,2}/d$ where the parameter $\delta _{1}=-0.19$ meV$\cdot $%
nm has been measured \cite{comm1} and the ratios $\delta _{0}/\delta
_{1}=4.5 $ and $\delta _{2}/\delta _{1}=-1.4$ can be estimated theoretically
\cite{Jeong09,Ando00,Brataas06}. The $\Delta _{0}$ depends on CNT chirality
so that $\Delta _{0}$ is maximum in zigzag CNTs ($\theta =0$) and minimum $%
(\Delta _{0}=0$) in armchair ones ($\theta =\pi /6$) \cite{Jeong09}.

These estimates show that the SOI may be treated as a perturbation for
actual electronic energies, i.e., $\left\vert \Delta _{i}\right\vert \ll
\left\vert \varepsilon _{k,m,\lambda }\right\vert $. In the first order, the
quantum averaging of the $H_{SO}$ on the eigenvectors $\left\vert
k,m,\lambda \right\rangle $ of Hamiltonian $H_{\lambda }$ gives rise to the
reduced Hamiltonian of SOI $H_{SO}^{\prime }$ of electron spin energy in an
effective field $\mathbf{B}_{SO}^{\prime }$ directed along the principal
axis $\zeta $ of the CNT, $H_{SO}^{\prime }=g\mu _{B}B_{SO}^{\prime
}S_{\zeta }$, where $g\mu _{B}B_{SO}^{\prime }=2\lambda (\Delta _{0}+\Delta
_{1}\sigma _{1})$. We take into account that $\left\langle k,m,\lambda
\left\vert \widehat{\sigma }_{0}\right\vert k,m,\lambda \right\rangle =1$, $%
\left\langle k,m,\lambda \left\vert \widehat{\sigma }_{2}e^{\pm i\varphi
}\right\vert k,m,\lambda \right\rangle =0$ and
\begin{equation}
\sigma _{1}=\left\langle k,m,\lambda \left\vert \widehat{\sigma }%
_{1}\right\vert k,m,\lambda \right\rangle =\mp \frac{\lambda
\varkappa _{m,\lambda }}{\sqrt{k^{2}+\varkappa _{m}^{2}}}.
\label{6d}
\end{equation}%
Here $\mp $ in Eq.~(\ref{6d}) corresponds to conduction (valence) band.

An important issue of Eq.~(\ref{6d}) is the dependence of $\mathbf{B}_{SO}^{\prime }(\sigma _{1})$ on $k$ i.e. electronic energy [eq.~(\ref{2b})]. This effect can be understood as a result of diminishing of electron velocity $ v_{\perp }$ in the $\rho $-direction with increasing $\ k_{\parallel }$ along the CNT axis under constant total electron velocity
$v_{F}$, i.e. $v_{\perp }/v_{F}=k_{\perp }/\left\vert
\mathbf{k}\right\vert =k_{\perp
}/\sqrt{k_{\perp }^{2}+k_{\parallel }^{2}}$ that appears as a factor in Eq.~(\ref{6d}) at $k_{\parallel }=k$ and $k_{\perp }=\varkappa _{m,\lambda }$.


It is convenient to introduce the mean values $\left\langle \sigma
_{1}\right\rangle =\sum\nolimits_{m}\int \sigma _{1}f(\varepsilon
_{k,m,\lambda })dk\diagup \sum\nolimits_{m}\int f(\varepsilon
_{k,m,\lambda })dk$ and $B_{SO}=\left\langle B_{SO}^{\prime
}\right\rangle =2\lambda (\Delta _{0}+\Delta _{1}\left\langle \sigma
_{1}\right\rangle )/g\mu _{B}$, where $f(\varepsilon _{k,m,\lambda
})$ is a thermal population factor for electron with energy
$\varepsilon _{k,m,\lambda }$. Correspondingly, the SOI can be
separated into a steady part $g\mu _{B}B_{SO}S_{\zeta }$ and a
fluctuating one $\Omega S_{\zeta }$ with%

\begin{equation}
\Omega =2\Delta _{1}\lambda \left( \sigma _{1}-\left\langle \sigma
_{1}\right\rangle \right) .  \label{7b}
\end{equation}%
Fluctuations in $k$ lead to fluctuations of the effective spin-orbital field
and serve as a mechanism of spin relaxation in CNTs. Our analysis centers on
the efficiency of this mechanism for CNTs in the presence of an external
magnetic field $\mathbf{B}_{0}$. This supplements the Hamiltonian $H_{SO}$
with a Zeeman interaction $H_{Z}=g\mu _{B}\mathbf{B}_{0}\mathbf{S}$. The
joint action determines a spin energy $g\mu _{B}\mathbf{B}_{eff}\mathbf{S}$
in an effective field $\mathbf{B}_{eff}=\mathbf{B}_{SO}+\mathbf{B}_{0}$.

In the reference frame ($\xi $, $\eta $, $\zeta $), the Zeeman Hamiltonian
takes the form $H_{Z}=\hbar \omega _{Z}(\sin \alpha \ S_{\xi }+\cos \alpha \
S_{\zeta })$, $\hbar \omega _{Z}=g\mu _{B}B_{0}$ ($g=2$ and $\mu _{B}$ is
the Bohr magneton) so that the frequency of spin precession $\omega
_{\lambda }=g\mu _{B}B_{eff}/\hbar $ in a total effective field $\mathbf{B}%
_{eff}$ is \cite{Churchill09}
\begin{equation}
\omega _{\lambda }=\sqrt{(\omega _{Z}\sin \alpha )^{2}+\left[ \omega
_{Z}\cos \alpha +2\lambda (\Delta _{0}+\Delta _{1}\left\langle \sigma
_{1}\right\rangle )/\hbar \right] ^{2}}.  \label{7d}
\end{equation}%
Spin relaxation is conveniently described in other co-ordinates $X$, $Y$ and
$Z$, where $Z$ is a spin quantization axis defined by the $\mathbf{B}_{eff}$%
. Correspondingly, evolution of $S_{Z}$ and $S_{X}$, $S_{Y}$ correlates with
longitudinal (spin-flip) and transversal (phase) spin relaxation. The angle $%
\alpha _{\lambda }$ ($0\leq \alpha _{\lambda }\leq \pi $) of $\mathbf{B}%
_{eff}$ slope to CNT axis defines%
\begin{equation}
\cos \alpha _{\lambda }=[\hbar \omega _{Z}\cos \alpha +2\lambda (\Delta
_{0}+\Delta _{1}\left\langle \sigma _{1}\right\rangle )]\diagup \hbar \omega
_{\lambda }.  \label{7f}
\end{equation}%
In a second quantization representation, the operator of fluctuating field
takes the form $\Omega =\sum\nolimits_{k,m}\mathbf{b}(k,m)n_{k,m}$, $%
n_{k,m}=a_{k,m}^{\dagger }a_{k,m}$, where $a_{k,m}^{\dagger }$ and $a_{k,m}$
are creation and annihilation operators and the vector $\mathbf{b}%
(k,m)=2\Delta _{1}\lambda \left( \sigma _{1}-\left\langle \sigma
_{1}\right\rangle \right) (-\sin \alpha _{\lambda },0,\cos \alpha _{\lambda
})$ is expressed in co-ordinates X, Y, and Z.

Assuming a short correlation time $\tau _{p}$ for the electronic correlation
function $\left\langle n_{k,m}(t)n_{k,m}\right\rangle $ in comparison with
the spin relaxation time $\tau _{S}$, one can use the Markovian equations of
spin evolution in form of expanded Bloch equations in which the spin mean
values $\left\langle \mathbf{S}\right\rangle =Tr\widehat{\rho }(t)\mathbf{S}$
($\widehat{\rho }(t)$ is a density matrix \cite{SSK}). Then the spin
polarization deviation from thermal equilibrium becomes $\mathbf{P}%
=2(\left\langle \mathbf{S}\right\rangle -\left\langle \mathbf{S}%
\right\rangle _{\infty })$ , where $2\left\langle \mathbf{S}\right\rangle
_{\infty }\mathbf{=}2\lim_{t\rightarrow \infty }\left\langle \mathbf{S}%
\right\rangle =\mathbf{(}0,0,-\tan \hbar \omega _{\lambda }/2k_{B}T\mathbf{)}
$. These equations take the simplest form in the $X$, $Y\,$,$\ Z$ coordinate
system:%
\begin{equation}
\frac{d\mathbf{P}_{\lambda }}{dt}=\left( \widehat{\mathbf{\omega }}_{\lambda
}-\widehat{\mathbf{\Gamma }}_{\lambda }\right) \mathbf{P}_{\lambda },
\label{eM}
\end{equation}%
where the $3\times 3$ matrix $\widehat{\mathbf{\omega }}_{\lambda }$
represents the cross product of $\overrightarrow{\mathbf{\omega }}_{\lambda }%
\mathbf{=(}0,0,\omega _{\lambda }\mathbf{)}$, $\widehat{\mathbf{\omega }}%
_{\lambda }\mathbf{P}_{\lambda }=\overrightarrow{\mathbf{\omega }}_{\lambda }%
\mathbf{\times P}_{\lambda }$. Thus, $\widehat{\mathbf{\omega }}_{\lambda }$
includes only two non-zero matrix elements $\omega _{YX}=-\omega
_{XY}=\omega _{\lambda }$,\ and $\widehat{\mathbf{\omega }}_{\lambda }%
\mathbf{S}_{0}=\mathbf{0}$. In this case the matrix of relaxation parameters
can be reduced to%
\begin{equation}
\widehat{\mathbf{\Gamma }}_{\lambda }\mathbf{=}\left(
\begin{array}{ccc}
R_{XX}^{(\lambda )} & 0 & R_{XZ}^{(\lambda )} \\
0 & R_{YY}^{(\lambda )} & 0 \\
R_{XZ}^{(\lambda )} & 0 & R_{ZZ}^{(\lambda )}%
\end{array}%
\right) .  \label{e3}
\end{equation}%
The low symmetry of the CNT affected by an arbitrary magnetic field
produces multiple spin-relaxation parameters $R_{\mu ,\nu
}^{(\lambda )}$, $\mu ,\nu =X$, $Y$, $Z$ {[}four in the case of
Eq.~(\ref{e3}){]}, which cannot be immediately associated with
longitudinal or transversal spin relaxation rates. In the most
general form they can be represented in terms of Fourier
transformation of correlation functions $\left\langle \Omega _{\mu
}(t)\Omega _{\nu }\right\rangle _{\omega }$ at frequency $\omega
=\omega _{\lambda }$. For spin precession in a fluctuating field
they can be evaluated in an approximation of a momentum relaxation
time $\tau _{k}$ \cite{Pikus,SSK} that assumes explicit knowledge
of the $\tau _{k}$ dependence on $k$. Instead, we use the average
value $\tau _{p}^{-1}=\left\langle \tau _{k}^{-1}\right\rangle $ to
evaluate the spin-relaxation characteristics while the specific
dependence of $\tau _{k}$ on $k$ can be taken into account by
introducing the numerical factor, which is about $Q_{1}\simeq
1.5\div 3$ \cite{Pikus}. Based on a recent
study \cite{Habenicht08}\ we assign $\tau _{p}=1$ ps and assume $%
Q_{1}=2$ for further numerical calculations.

Applying these approximations we find
\begin{eqnarray}
R_{XX}^{(\lambda )} &=&\cos ^{2}\alpha _{\lambda }\frac{(2\Delta
_{1})^{2}\tau _{p}}{\hbar ^{2}}Q_{1}\Delta \sigma _{1}^{2};  \nonumber \\
R_{ZZ}^{(\lambda )} &=&\sin ^{2}\alpha _{\lambda }\frac{(2\Delta
_{1})^{2}\tau _{p}}{\hbar ^{2}(\omega _{S}^{2}\tau _{p}^{2}+1)}Q_{1}\Delta
\sigma _{1}^{2};  \label{e5} \\
R_{YY}^{(\lambda )} &=&R_{XX}^{(\lambda )}+R_{ZZ}^{(\lambda )};  \nonumber \\
R_{XZ}^{(\lambda )} &=&\frac{R_{XX}^{(\lambda )}}{2}\tan \alpha _{\lambda }+%
\frac{R_{ZZ}^{(\lambda )}}{2}\cot \alpha _{\lambda },  \nonumber
\end{eqnarray}%
where $\Delta \sigma _{1}^{2}=\left\langle \sigma
_{1}^{2}\right\rangle -\left\langle \sigma _{1}\right\rangle ^{2}$
critically depends on the temperature so that $\Delta \sigma
_{1}\rightarrow 0$ at $T\rightarrow 0$. Results of calculations of
the relaxation rate parameters [Eq.~(\ref{e5}]) as a function of
temperature, magnetic field strength and direction, and CNT diameter
are presented in Fig.~2. It can be noticed that there is not a
single $R_{\mu \nu }^{(\lambda )}$, which significantly dominates
over all ranges of $T$, $\mathbf{B}_{0}$, and $d$. Thus the full set
of $R_{\mu \nu }^{(\lambda )}$ is required to describe spin
relaxation in CNTs. All $R_{\mu \nu }^{(\lambda )}$ are strongly
depended on temperature that controls energy dispersion. Moreover,
the relaxation rate decreases with an increase of CNT diameter due
to SOI reduction with decreasing curvature. The non-monotonic
dependence on the external magnetic
field strength and its direction depends upon the interplay between $\mathbf{%
B}_{SO}$ and $\mathbf{B}_{0}$ that leads to suppression of $R_{ZZ}^{(\lambda
)}$ and $R_{XZ}^{(\lambda )}$ when $\alpha \rightarrow 0$ or $%
B_{0}\rightarrow 0$.

Examples of numerical solutions to Eqs.~(\ref{eM},\ref{e3},\ref{e5})
are shown in Fig.~3. In particular, interference of spin
polarizations from the two non-equivalent valleys can lead to the
oscillation amplitude beating shown in Fig. 3(a) where beats
alternate with nodes in ~ 90 ps. When the magnetic field is
perpendicular to the CNT axis ($\alpha =90^{\circ }$), a fast
polarization damping is found for longitudinal ($\tau _{s}\simeq
150$ ps) and transversal ($\tau _{s}\simeq 110$ ps) spin
polarizations as seen in Fig. 3(b). A greater variation of spin
relaxation exists in CNTs with different chirality placed in an
arbitrary directed magnetic field.

The developed theory can be applied to MR measurements which are usually
carried out on a CNT with weak contacts to source/drain ferromagnets.
Electron undergo multiple reflections from the contacts before exiting. In
such a case the output signal depends on both spin relaxation time $\tau _{s}
$ and electron dwell time $\tau _{d}$ \cite{Hueso07}. The experimental setup
also provides for electron injection into the CNT with spin polarization
along magnetization $\mathbf{M}_{S}$\ of ferromagnetic source and spin
detection with a ferromagnetic drain polarized in either the same direction $%
\mathbf{M}_{D}=\mathbf{M}_{S}$ or the opposite one, $\mathbf{M}_{D}^{\prime
}=-\mathbf{M}_{S}$. Spin dependent output from each valley $\Delta
I_{\lambda }$ is proportional to difference $\mathbf{m}\overline{\mathbf{P}}%
_{\lambda }-\mathbf{m}^{\prime }\overline{\mathbf{P}}_{\lambda }=2\mathbf{m}%
\overline{\mathbf{P}}_{\lambda }$ ($\mathbf{m}=\mathbf{M}_{D}/\left\vert
\mathbf{M}_{D}\right\vert $, $\mathbf{m}^{\prime }=-\mathbf{m}$) where the
spin polarization is averaged over $\tau _{d}$ ( $\overline{\mathbf{P}}%
_{\lambda }=\frac{1}{\tau _{d}}\int\nolimits_{0}^{\infty }\mathbf{P}%
_{\lambda }(t)e^{-t/\tau _{d}}dt$) provided that $\tau _{d}$ is longer than
electron drift from source to drain. $\overline{\mathbf{P}}_{\lambda }$ can
be found from Eq.~(\ref{eM}) by multiplication with $e^{-t/\tau _{d}}$ and
subsequent integration over $t$. Assuming that the rate of intervalley
transitions is less than $R_{d}=1/\tau _{d}$ , the total spin-depended
output can be expressed as $\sum\nolimits_{\lambda }\Delta I_{\lambda
}/2=F_{s}R_{d}$, where
\begin{equation}
F_{s}=c\mathbf{m}\sum\nolimits_{\lambda }(\widehat{\mathbf{\Gamma }}%
_{\lambda }+\widehat{\mathbf{I}}R_{d}-\widehat{\mathbf{\omega }}_{\lambda
})^{-1}\mathbf{m}.  \label{eR}
\end{equation}%
Here\ $c$ is a dimensional constant and $\widehat{\mathbf{I}}$ is the
identity $3\times 3$ matrix. In the case of $\omega _{\lambda }=0,$ $%
\widehat{\mathbf{\Gamma }}_{\lambda }=\widehat{\mathbf{I}}/\tau _{s}$, Eq.~(%
\ref{eR}) reduces to $F_{s}=c/(\tau _{d}^{-1}+\tau _{s}^{-1})$ which
is in agreement with previous calculations \cite{Hueso07}. In more
general situations, the following relation holds%
\begin{equation}
\tau _{s}=-\lim_{R_{d}\rightarrow 0}\frac{d\ln F_{s}}{dR_{d}}.  \label{a1}
\end{equation}%
If $\omega _{\lambda }\gg R_{\mu \nu }^{(\lambda )}\equiv 1/\tau
_{\mu \nu }^{(\lambda )}$, the last equation leads to $\tau
_{s}=\sum\nolimits_{\lambda }\tau _{ZZ}^{(\lambda )2}\diagup
\sum\nolimits_{\lambda }\tau _{ZZ}^{(\lambda )}$. Thus, only
longitudinal spin relaxation (i.e. spin-flip) can control the
magnetoresistance measurements in a CNT at strong magnetic field. In
Fig. 3c the calculated dependence of $\tau _{s}$ on the chiral angle
$\theta $ of the CNT is displayed. For example, $\tau _{s}\simeq
220$ ps at $T=$300 K, for
the case of $B_{0}=$0.5 T, $d=$2 nm, $\mathbf{m}\parallel \mathbf{B}_{0}$ $%
\alpha =90^{\circ }$ and $\theta =26^{\circ }$.

However spin relaxation is suppressed at small values of $\alpha $.
The experimental setup treated in Ref. \cite{Hueso07} displays a
small deviation of the magnetization direction from the CNT axis,
with $\alpha
\simeq 10^{\circ }$. For such a case and with $T=120$ K, $B_{0}=$0.1 T and $%
d=$ 10 nm Eq.~(\ref{a1}) predicts $\tau _{s}\simeq 69$ ns which
approaches the measured $\tau _{s}=$30 ns. This results indicates
the high efficiency of the spin relaxation mechanism even through
small symmetry breaking in the design of realistic devices.

In conclusion, we show that the fluctuations of curvature-induced
spin-orbital interaction that are associated with electron random
motion in CNTs is responsible for the short spin relaxation. If a
magnetic field preserves the uniaxial symmetry of CNT, the spin-flip
relaxation is suppressed. However a small asymmetry may lead to
visible manifestations of this precession change. These issues may
stimulate further experiments to determine the optimal conditions
for room temperature CNT spintronic device applications.


This work was supported in part by the US Army Research Office, NSF, and the
FCRP Center on Functional Engineered Nano Architectonics (FENA).

\clearpage
\begin{figure}[tbp]
\includegraphics[scale=.6,angle=0]{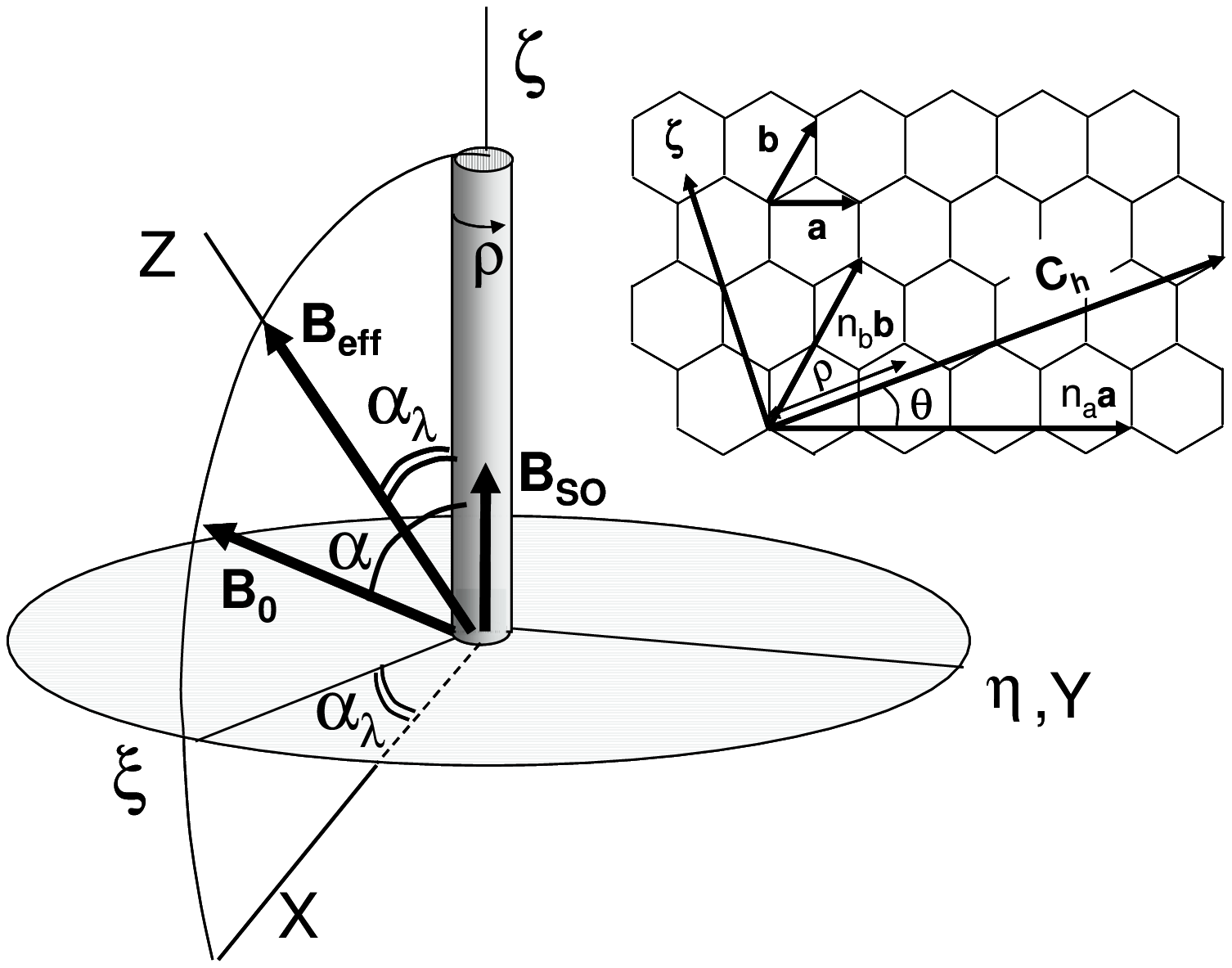}
\caption{Reciprocal positions of CNT and coordinate systems. Insert:
the lattice structure of graphene.}
\end{figure}

\clearpage
\begin{figure}[tbp]
\includegraphics[scale=.6,angle=0]{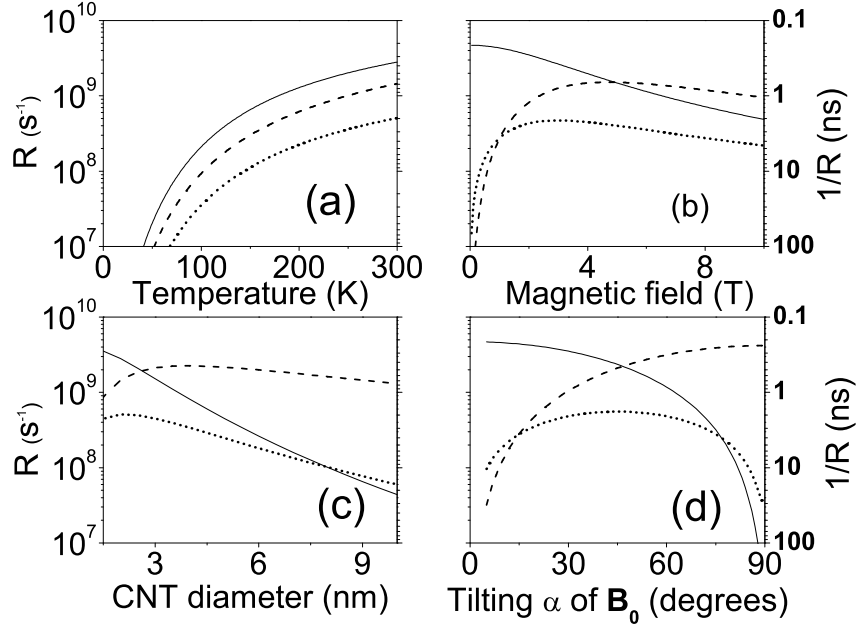}
\caption{$R_{XX}$ (solid lines), $R_{ZZ}$ (dashed lines) and
$R_{XZ}$ (doted lines) calculated as an average over valleys and the
functions of (a)
temperature {[}$(n_{a},n_{b})=(15,14)$, $B_{0}=2$ T, $\protect\alpha=\protect%
\pi/2${]}, (b) magnetic field strength {[}$(n_{a},n_{b})=(15,14)$,
$T=300$ K, $\protect\alpha=\protect\pi/2$ {]}, (c) the nanotube
diameter {[}$T=300$
K, $B_{0}=2$ T, $\protect\alpha=\protect\pi/2$, $\protect\theta=27^{\circ}${]%
} and (d) the angle $\protect\alpha$ between external magnetic field
direction and CNT axis {[}$(n_{a},n_{b})=(17,13)$, $T=300$ K, $B_{0}=2$ T{]}%
. For convenience the right parts of the graphs are scaled in
nanoseconds as a reciprocal values.}
\end{figure}

\clearpage
\begin{figure}[tbp]
\includegraphics[scale=.9,angle=0]{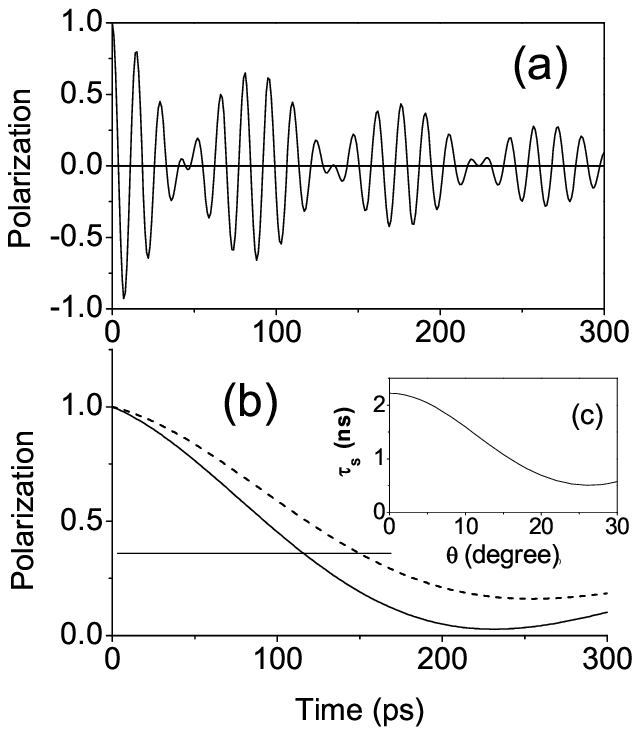}
\caption{(a) $(n_{a},n_{b})=(15,14)$, $B_{0}=0.2$ T and
$\protect\alpha=30^{\circ}$ (b) $(n_{a},n_{b})=(17,13)$, $
B_{0}=0.05$ T and $\protect\alpha=90^{\circ}$. Solid (dashed) lines
corresponds to transversal $P_{\protect\xi}$ (longitudinal $P_{\protect\zeta%
} $) spin polarization. Thin horizontal line cuts the polarization
curves at the level $1/e$, which indicates the relaxation times 110
ps and 150 ps for $P_{\protect\xi}$ and $P_{\protect\zeta}$. Insert
(c) displays the dependence of relaxation time $\protect\tau_{S}$ on
chiral angle calculated with Eq.~(\protect\ref{a1}) under constant
$d=10$ nm, $B_{0}=2$ T and $\protect\alpha=90^{\circ}$.}
\end{figure}


\begin{thebibliography}{99}
\bibitem{comm1} Evaluating the $\Delta_{1}$ in Ref. \cite{Kuemmeth08}, the
contribution of first term $\sim\Delta_{0}$ in Eq.~(\ref{3}) was not
taken into account. It does not matter if $\cos3\theta\approx0$; we
suppose that this was the case for the actual CNT in
Ref.~\onlinecite{Kuemmeth08}.
\end{thebibliography}
\end{document}